\documentclass[sigconf]{acmart}
\AtBeginDocument{%
  }

\usepackage{enumitem}
\usepackage{calc} 
\usepackage{balance}
\usepackage{graphicx}
\usepackage{subcaption}
\usepackage{multirow}

\copyrightyear{2025}
\acmYear{2025}
\setcopyright{acmlicensed}\acmConference[SIGIR '25]{Proceedings of the 48th International ACM SIGIR Conference on Research and Development in Information Retrieval}{July 13--18, 2025}{Padua, Italy}
\acmBooktitle{Proceedings of the 48th International ACM SIGIR Conference on Research and Development in Information Retrieval (SIGIR '25), July 13--18, 2025, Padua, Italy}
\acmDOI{10.1145/3726302.3730159}
\acmISBN{979-8-4007-1592-1/2025/07}

\begin{document}

\title{A Human-AI Comparative Analysis of Prompt Sensitivity in LLM-Based Relevance Judgment}

\author{Negar Arabzadeh}
\email{narabzad@uwaterloo.ca}
\orcid{0000-0002-4411-7089}
\affiliation{%
  \institution{University of Waterloo}
  \city{Waterloo}
  \state{Ontario}
  \country{Canada}
}

\author{Charles L.A. Clarke}
\email{claclark@uwaterloo.ca}
\orcid{0000-0001-8178-9194}
\affiliation{%
  \institution{University of Waterloo}
  \city{Waterloo}
  \state{Ontario}
  \country{Canada}
}

\begin{abstract}

Large Language Models (LLMs) are increasingly used to automate relevance judgments for information retrieval (IR) tasks, often demonstrating agreement with human labels that approaches inter-human agreement.
To assess the robustness and reliability of LLM-based relevance judgments, we 
systematically investigate impact of prompt sensitivity on the task.
We collected prompts for relevance assessment from 15 human experts and 15 LLMs across three tasks~---~binary, graded, and pairwise~---~yielding 90 prompts in total.
After filtering out unusable prompts from three humans and three LLMs, we employed the remaining 72 prompts with three different LLMs as judges to label document/query pairs from two TREC Deep Learning Datasets (2020 and 2021).
We compare LLM-generated labels with TREC official human labels using Cohen's $\kappa$ and pairwise agreement measures.
In addition to investigating the impact of prompt variations on agreement with human labels, we compare human- and LLM-generated prompts and analyze differences among different LLMs as judges.
We also compare human- and LLM-generated prompts with the standard UMBRELA prompt used for relevance assessment by Bing and TREC 2024 Retrieval Augmented Generation (RAG) Track.
To support future research in LLM-based evaluation, we release all data and prompts at \url{https://github.com/Narabzad/prompt-sensitivity-relevance-judgements/}.

\end{abstract}


\maketitle

\section{Introduction}
Large Language Models (LLMs) are increasingly used for evaluation across various domains, including natural language processing and automated content assessment \cite{alaofi2024generative,rahmani2024LLM4eval,salemi2024evaluating,chang2024survey,chiang2023can,arabzadeh-etal-2024-assessing}. 
The information retrieval (IR) community has been an early adopter of LLMs for relevance assessment~\cite{faggioli2023perspectives,thomas2023large,li2024generation,zhuang2023beyond,meng2024query}.
Numerous studies have confirmed that LLM-generated relevance labels
closely align with human labels under multiple measures of agreement~\cite{upadhyay2024umbrela,upadhyay2024largescalestudyrelevanceassessments,macavaney2023one}.

Nonetheless, despite the widespread adoption of LLMs for relevance assessment, prompting strategies vary substantially across studies~\cite{farzi2024pencils,sander2021exam,arabzadeh2024adapting,arabzadeh2024comparison}.
An experiment reported at the LLM4Eval Workshop in SIGIR 2024 on Large Language Models for Evaluation in Information Retrieval \cite{rahmani2024report1stworkshoplarge},  analyzed how different prompts influence agreement with human judgments and system rankings~\cite{rahmani2024LLM4eval}.
While multiple studies have examined how LLMs respond to different prompting strategies~\cite{sclar2023quantifying,chatterjee2024posix,lu-etal-2024-prompts,arora2022askanythingsimplestrategy,leidinger2023languagepromptinglinguisticproperties}, these studies have generally been conducted with prompts tuned to specific LLMs and collections, or where prompt variants are constrained by templates \cite{azzopardi2024report}.
As a complement to these studies, we report on a study of prompts from a variety of independent sources that have not been tuned to LLMs or collections, allowing us to examine the robustness of LLM-based relevance assessment under different prompting strategies.
This investigation also allows us to compare different LLMs as judges to determine the degree to which different LLMs are sensitive to prompt modifications.

We collected and analyzed prompts generated by both human experts and LLMs themselves.
We designed a guideline for prompting LLMs to perform relevance assessment following three different approaches: \textit{binary}, \textit{graded}, and \textit{pairwise}.
While most previous studies have focused on graded relevance, we believe it is crucial to explore a wider range of relevance assessment methods, as they have proven effective in assessing different scenarios in the evaluation of information-seeking systems \cite{cleverdon1991significance,10.1145/3451161,Yan_2022,kyct13,cvs21,cbcd08,sz20,xie20,buckley2004retrieval,hawking1999overview,voorhees2000report}.
As a benefit to employing LLMs for relevance assessment, it becomes easier to explore different approaches to relevance assessment since human judges do not need to be recruited and trained separately for each approach.

We recruited {15 human participants} to create prompts for each of the three assessment approaches.
As part of the recruitment process, we ensured that the participants were familiar with prompt engineering and relevance assessment principles, as detailed in Section \ref{sec:data}.
As a result of this inclusion criteria for recruitment, most participants were drawn from three academia NLP/IR labs.
We also collected prompts from 15~different open source and commercial LLMs. 
Our primary goal is to understand prompt sensitivity in LLM-based relevance judgment \cite{razavi2025benchmarking}, including its impact, robustness, and variation across different LLMs.
Additionally, we explore the effectiveness of LLM as prompt generators.


We performed relevance judgment experiments using data from two years of the TREC Deep Learning Track: DL~2020~\cite{trecdl2020}, and DL~2021~\cite{trecdl2021}.
Using the prompts created by both human participants and LLMs, we conducted relevance assessments on query-document pairs from these datasets using two open-source LLMs~---~\texttt{LLaMA 3.2-3b} and \texttt{Mistral 7b}~---~and one commercial LLM \texttt{GPT-4o}. 
Our experiment incorporates the three approaches to relevance assessment (binary, graded, and pairwise) with prompts from both humans and LLMs using {three different LLMs as judges}.
Through our experiments, we address the following research questions:

\begin{itemize}[leftmargin=*]
\item  \textbf{RQ1. Impact of Prompts on LLM-based Relevance Judgment Approaches:}  
 Given a clear {task objective}, how do different prompts influence the effectiveness of each approach to LLM-based relevance judgment?
 \item \textbf{RQ2. LLMs as Prompt Generators:}   How effective are LLM-generated prompts for relevance judgment, and how do they compare to human-crafted prompts?
\item \textbf{RQ3. Prompt Robustness Across LLMs:} Are there prompts that consistently perform well across different LLMs, regardless of the model used as a judge?
\item \textbf{RQ4. Model-Specific Sensitivity to Prompts:}  Is prompt sensitivity consistent across all models, or do some LLMs show greater variability in performance? 
\end{itemize}
To ensure {reproducibility}, we have made all {data and experimental artifacts} publicly available at \url{https://github.com/Narabzad/prompt-sensitivity-relevance-judgements/}.
The study reported in this paper, and its associated data release, has received ethics clearance as human subjects research from our institution. 

\section{Prompt Creation}
\label{sec:data}

\subsection{Prompt generation}
To investigate the impact of prompting on LLM-based relevance judgment, we collected data from both human participants and LLMs, ensuring that the task objective remained clear and consistent (sharing the same intent) across all participants. 
We prepared guidelines for prompt writing\footnote{\url{https://bit.ly/4hP0EMg}}, which provides detailed explanations of the three relevance judgment tasks:
1) Binary relevance~---~a passage is either relevant (1) or not relevant (0) to a query.
2) Graded relevance~---~a passage is rated on a 0-3 scale, where 3 indicates perfect relevance to the query.
3)  Pairwise relevance~---~given two passages, chose the passage more relevant to the query.
In the guideline, 
each task is illustrated with examples from the TREC Deep Learning 2019 \cite{trecdl2019}, helping to ensure that both humans and LLMs had a well-defined understanding of the task.
These examples could also be used as (few shot) examples if desired.

The guidelines specify a Python-based format, where participants (both human and LLMs) were required to fill in structured Python dictionaries.
More specifically, participants had to provide both the \texttt{"system message"} and \texttt{"user message"} fields for the prompts, following the format commonly used in LLM-based prompting (e.g., OpenAI models and open-source alternatives such as those from Ollama). This structured approach ensures compatibility across different LLM implementations.

We recruited 15 human participants, each of whom had at least a Master’s degree in computer science, were fluent in English, and had prior experience working with LLMs via API usage or coding.
Additionally, these participants had previously published at least one paper in an IR-focused conference.
Each participant received a \$10 gift card as a token of appreciation for their time and effort.

For prompt creation, we also used 15 different LLMs from the ChatBotArena\footnote{\url{https://lmarena.ai/}} platform \cite{chiang2024chatbot}, which enables the execution of various LLMs online.
We provided the same data collection guideline to the LMMs, including the task description and examples, ensuring that the LLMs received identical instructions to those given to human participants.
Similar to human participants, each LLM was asked to complete the \texttt{"system message"} and \texttt{"user message"} fields in our Python function for relevance judgment. This setup allow us to systematically compare the impact of prompting across both groups. Table \ref{tab:LLMs} provides the list of LLMs we used in this experiment for generating prompts for relevance judgments. 

\begin{table}[]
    \centering
    \caption{List of LLMs used for prompt generation.}
    \label{tab:LLMs}
    \scalebox{0.75}{
    \begin{tabular}{lllll}
        \toprule
        \midrule
        \texttt{GPT-4o} & \texttt{GPT-4o Mini} & \texttt{Claude 3.5} & \texttt{LLaMA 3.2} & \texttt{Phi-4} \\
        \texttt{Mistral-large} & \texttt{DeepSeek-v3} & \texttt{Amazon-Nova-Pro-v1} & \texttt{Gemma-2-9b} & \texttt{Grok-2} \\
        \texttt{Gemini 2} & \texttt{Jamba-1.5} & \texttt{Athene-v2} & \texttt{GPTO1} & \texttt{GPTO1 Mini} \\
        \bottomrule
    \end{tabular}}
\end{table}

\subsection{Filtering and cleaning}
To maintain consistency, we did not modify or provide additional instructions for any LLMs or human participants.
Among the LLMs, two failed to complete the task because they deemed the task to be inappropriate, or repeatedly asked about examples. 
Among human participants
, only one used a few-shot approach with examples.
The rest did not provide any examples in their prompts.
When testing the outputs of the collected prompts, not all of them were able to generate the expected format cleanly. 
Some prompts produced responses that required additional cleaning, such as verbose outputs like \textit{"The passage is relevant, so the answer is: 1"} instead of simply returning \texttt{1}.
To ensure consistency, we examined the all generated output and applied necessary cleaning.
After filtering and cleaning, we finalized 12 human-generated prompts and 12 LLM-generated prompts for use in our experiments.

\subsection{Prompt Diversity}

\begin{figure}[t]
    \centering
    \includegraphics[width=0.8\linewidth, trim={0.2cm 0.8cm 0.1cm 0.5cm}, clip]{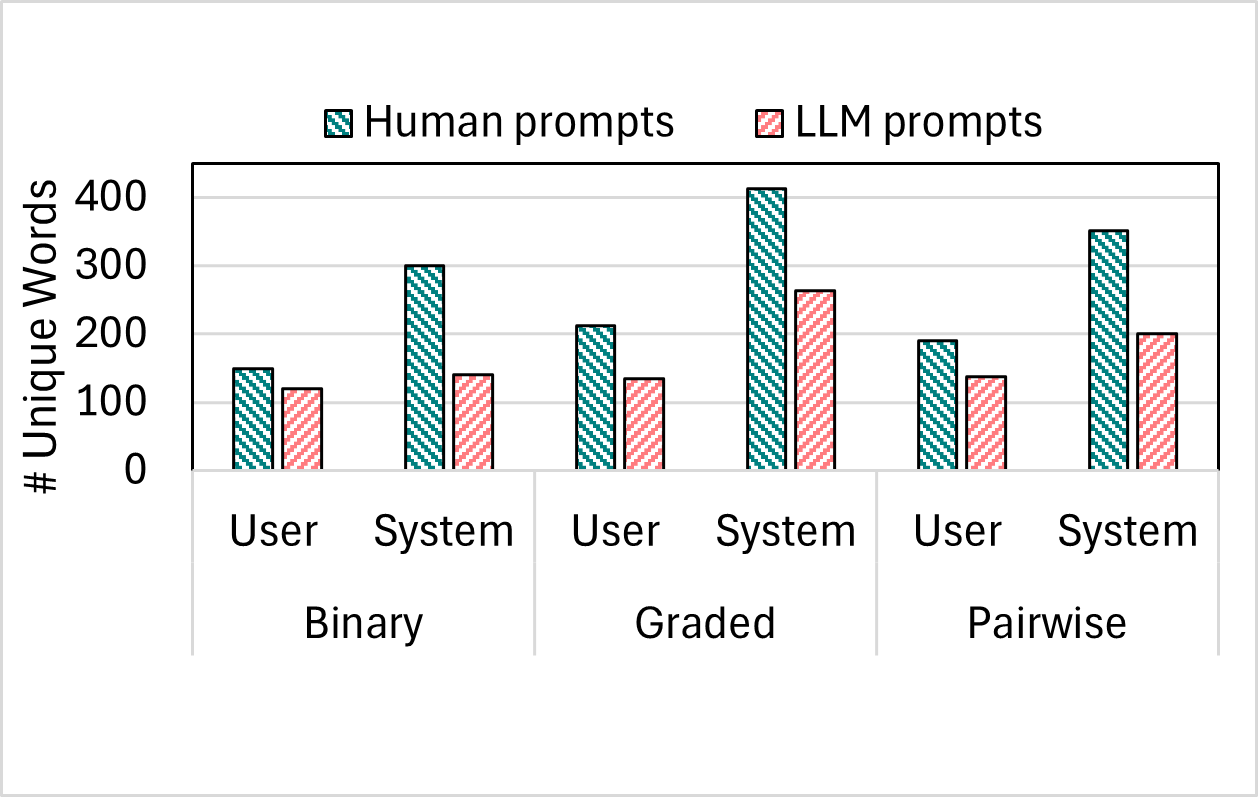}
    \caption{Diversity of words across human and LLM-generated prompts.}
    \label{fig:prompt_diversity}
\end{figure}

To better understand the variation in prompts, we examined the diversity of both human-generated and LLM-generated prompts. Specifically, we analyzed both \textit{user prompts} and \textit{system prompts} separately, as they serve distinct roles in guiding the LLM's response.
In a prompt the user message provides the direct instructions given to the model, specifying what information is needed. 
In contrast, the system message provides context for the task, defining the LLM's role and expected behavior (e.g., ``You are an expert relevance judgment assessor''). 
Figure~\ref{fig:prompt_diversity} illustrates the distribution of unique terms used across all human-generated (in green) and LLM-generated (in red) prompts. As shown in this figure, human-generated prompts exhibit greater diversity in wording when compared to LLM-generated ones. 
This suggests that humans introduce more nuanced descriptions and varied phrasing when defining the task, while LLM-generated system prompts tend to rely on more standardized language.
Additionally, system messages exhibit greater lexical diversity compared to user messages.

\section{Experimental Methodology}

\begin{figure*}[t]
    \centering
    \includegraphics[width=0.97\linewidth, trim={0cm 0cm 0cm 0cm}, clip]{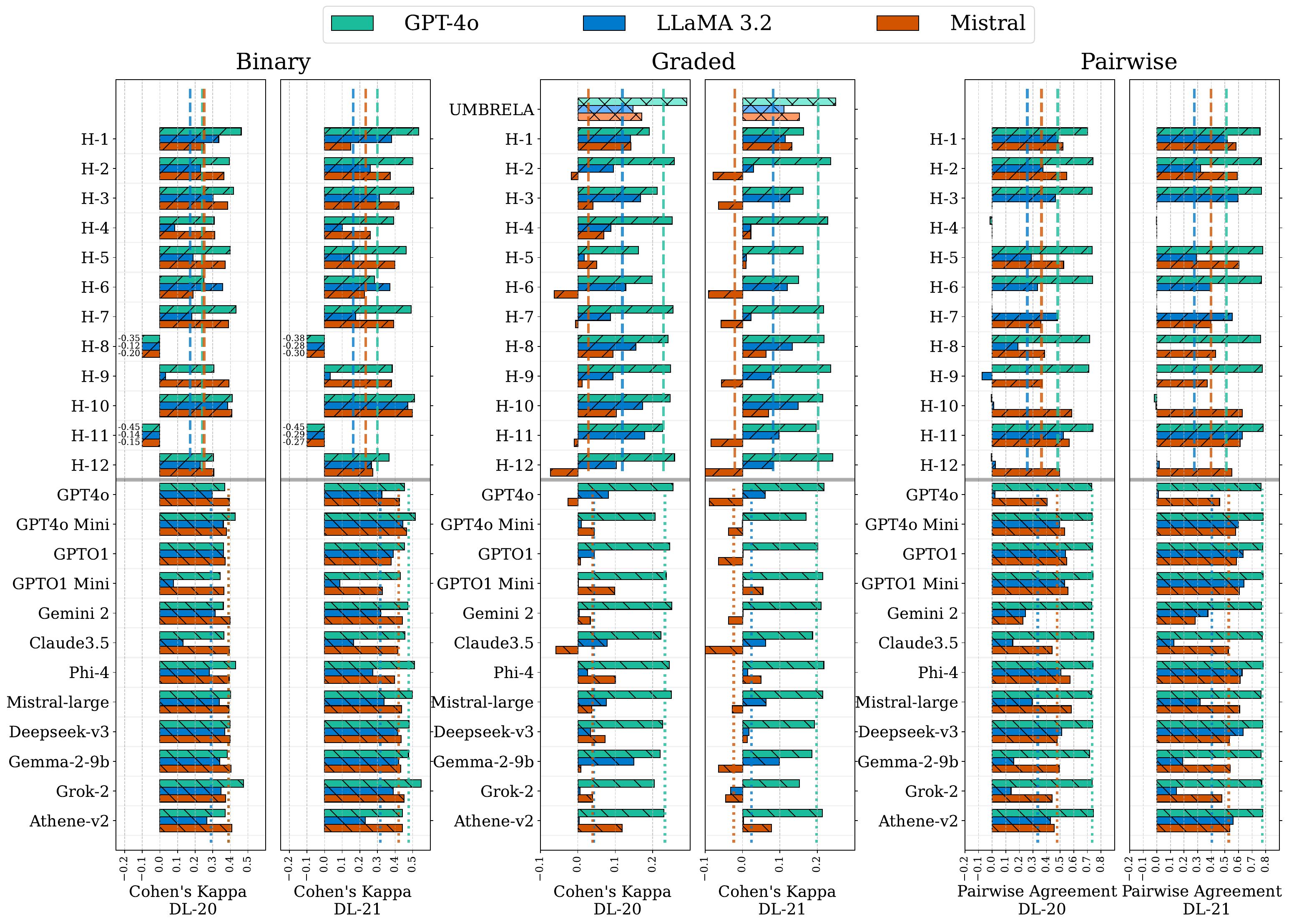}
    \caption{Agreement of LLM-based relevance judgments with human annotations across different prompts and relevance judgment tasks. UMBRELA represents the reproduction of Bing's LLM assessor introduced in \cite{upadhyay2024umbrela}. Otherwise, the top 12 bars (H-*) represent human-crafted prompts, while the bottom 12 correspond to LLM-generated prompts. The dashed lines show the mean of agreement in LLM    -crafted prompts and human-crafted prompts separately.}
    \label{fig:cohen}
\end{figure*}

\textit{Data} We utilize the TREC Deep Learning Track datasets from 2020 and 2021. The DL-20 dataset contains 54 judged queries with 11,386 relevance assessments from  MS MARCO V1 collection, while the DL-21 dataset includes 53 judged queries and 10,828 assessments from MS MARCO V2.
Both datasets have been manually annotated by NIST assessors following the TREC relevance judgment guidelines. The assessors evaluate each document-query pair based on a graded relevance scale, ranging from not relevant (0) to highly relevant (3). The assessment process involves pooling top-ranked documents from multiple retrieval systems, which were then judged by human annotators. Using this data allows us to compare the three different variations of LLM-based  judgments i.e., binary, graded, and pairwise. For graded relevance, we compare against the actual graded labels. For binary judgments, following prior work \cite{upadhyay2024umbrela,faggioli2023perspectives}, we classify levels 2 and 3 as relevant and levels 0 and 1 as non-relevant. For pairwise judgments, we compare documents with different relevance levels, assuming that a document with a higher relevance level should be ranked as more relevant than one with a lower relevance level.

\textit{LLMs for Relevance Judgments.} To perform relevance assessment, we employed three different LLMs: one commercial model, \texttt{GPT-4o}, and two open-source models, \texttt{LLaMA 3.2-3B} and \texttt{Mistral-7B} . We implemented our experiments using OpenAI and Ollama, running all prompts with a temperature setting of 0.

\textit{Data Sampling.} We conducted experiments on all query-document pairs for binary and graded relevance judgments using the open-source models. However, due to computational constraints, we were unable to run all 24 valid prompts across all query-document pairs for \texttt{GPT-4o}. Instead, we randomly sampled up to 10 documents per query for each of the four relevance levels (0-3). If fewer than 10 documents were available for a given relevance level, we included all available documents. 
For pairwise judgments, evaluating all possible pairs was not feasible due to their quadratic growth. Instead, 
we categorized documents for each query into three groups: ``highly relevant'', ``relevant'', and ``non-relevant''. The ``highly relevant'' category corresponds to the highest available relevance level for that query, which in TREC-style annotations could be level 3 or level 2, depending on availability. The ``non-relevant'' category includes all level 0 documents, while any intermediate relevance level (typically level 1, or levels 1 and 2 if level 3 exists) was classified as ``relevant''.

From these three categories, we constructed document pairs for pairwise judgments. Specifically, we sampled 10 pairs per query from each of the following comparisons: ``highly relevant vs.\ non-relevant'', ``relevant vs.\ non-relevant'', and ``highly relevant vs.\ relevant'' (up to 30 pairs in total).
If fewer than 10 pairs were available for a given comparison, we included as many as possible.
Additionally, for the pairwise setting, we minimized positional bias by evaluating each document pair twice, swapping the order of the documents in the second run.
The result is counted as ``agree'' if the LLM favors the more relevant passage in both comparisons, ``tie'' if the LLM's decisions are inconsistent when the passage order is swapped, and ``disagree'' if the LLM consistently selects the passage with a lower relevance level assigned by human annotators.

\begin{table}[t]
    \centering
    \caption{Mean and variance of agreement between LLM-based and human relevance judgments across different settings.}
    \label{tab:agreement_stats}
    \scalebox{0.8}{
    \begin{tabular}{llcccccc}
        \toprule \hline
        \multirow{2}{*}{Model} & \multirow{2}{*}{crafted by} & \multicolumn{2}{c}{Binary} & \multicolumn{2}{c}{Graded} & \multicolumn{2}{c}{Pairwise} \\
        \cmidrule(lr){3-4} \cmidrule(lr){5-6} \cmidrule(lr){7-8}
        & & Mean & Variance & Mean & Variance & Mean & Variance \\
        \midrule
        \multirow{2}{*}{\texttt{GPT-4o}} & LLM & 0.434 & 0.003 & 0.215 & 0.001 & 0.849 & 0.000 \\
         & Human & 0.270 & 0.098 & 0.215 & 0.001 & 0.578 & 0.139 \\
        \midrule
        \multirow{2}{*}{\texttt{LLaMA 3.2}} & LLM & 0.303 & 0.010 & 0.033 & 0.002 & 0.439 & 0.066 \\
         & Human & 0.167 & 0.041 & 0.102 & 0.003 & 0.330 & 0.073 \\
        \midrule
        \multirow{2}{*}{\texttt{Mistral}} & LLM & 0.405 & 0.001 & 0.008 & 0.004 & 0.574 & 0.014 \\
         & Human & 0.243 & 0.051 & 0.004 & 0.005 & 0.442 & 0.073 \\
        \bottomrule \hline
    \end{tabular}}
\end{table}
\section{Results and Findings}
In order to explore the research questions raised in the introduction, we investigated the agreement of LLM-based relevance judgments from different prompts with human annotations on TREC 2020 and 2021 using three different LLMs, as shown in Figure~\ref{fig:cohen}. 
For binary and graded relevance judgments, agreement is measured using Cohen's Kappa ($\kappa$). For pairwise judgments, since the task involves assessing agreement with the actual ranking of pairs, we report the percentage of cases where the LLM's preference agrees with the expected order.
In this figure, the leftmost two columns represent the results for binary, the middle two columns correspond to graded, and the rightmost two columns display the results from pairwise relevance judgment. 
The green, blue, and red bars indicate agreement for \texttt{GPT-4o}, \texttt{LLAMA 3.2}, and \texttt{Mistral}, respectively.
In each pair of plots, the left plot presents results for DL-20, while the right plot corresponds to DL-21.
The bottom 12 bars represent prompts crafted by LLMs; on top of them there are 12 bars corresponding to prompts created by humans.

In addition to results from the human- and LLM-written prompts, we also report the results of UMBRELA assessments at the top of the graded relevance sub-figure (middle). UMBRELA is an open-source reproduction of Microsoft's Bing LLM-based relevance assessor \cite{thomas2023large}, designed to automate relevance judgments effectively \cite{upadhyay2024umbrela,upadhyay2024largescalestudyrelevanceassessments}. It follows a structured prompting approach and has demonstrated high correlation with both human annotations and system rankings across multiple TREC Deep Learning Tracks (2019–2023). Notably, UMBRELA has been integrated into TREC 2024 RAG for automated evaluation, which further validated its reliability as an alternative to human assessors. We consider UMBRELA a reliable and effective prompt and we believe comparing its performance against human-crafted and LLM-generated prompts in graded relevance judgments would bring interesting insights.
Additionally, Table \ref{tab:agreement_stats} summarizes Figure \ref{fig:cohen} by providing the mean and variance of agreement scores across the two datasets and different relevance judgments.

We now consider investigating each of our research questions in light of these agreement results.

\textbf{RQ1. Impact of Prompts on LLM-based Relevance Judgment Approaches:}  
Figure~\ref{fig:cohen} and Table~\ref{tab:agreement_stats} reveal significant variance across different LLM-based relevance judgment approaches. Binary and pairwise methods exhibit the least sensitivity to input prompts, maintaining more consistent agreement. In contrast, graded relevance judgments are highly sensitive to prompt variations. We note that while binary and pairwise methods operate with only two choices, graded relevance introduces greater variability. Particularly on graded judgments, \texttt{GPT-4o} demonstrates relatively stable performance but \texttt{LLaMA 3.2} and \texttt{Mistral} show considerable fluctuations across different prompts.
 
\textbf{RQ2. LLMs as Prompt Generators:} 
Table~\ref{tab:agreement_stats} shows that LLM-generated prompts generally yield higher average agreement with human annotations. However, for graded relevance judgments, the difference is minimal. This may be due to (i) participants’ greater familiarity with graded assessments or (ii) the inherently subjective nature of assigning relevance levels, which may require more calibration with human annotators. Additionally, LLM-generated prompts exhibit lower variance in agreement compared to human-crafted prompts, indicating less sensitivity to prompt variations.

\begin{figure}[t]
    \centering
    \includegraphics[width=0.8\linewidth, trim={0.2cm 0.2cm 0.1cm 0.1cm}, clip]{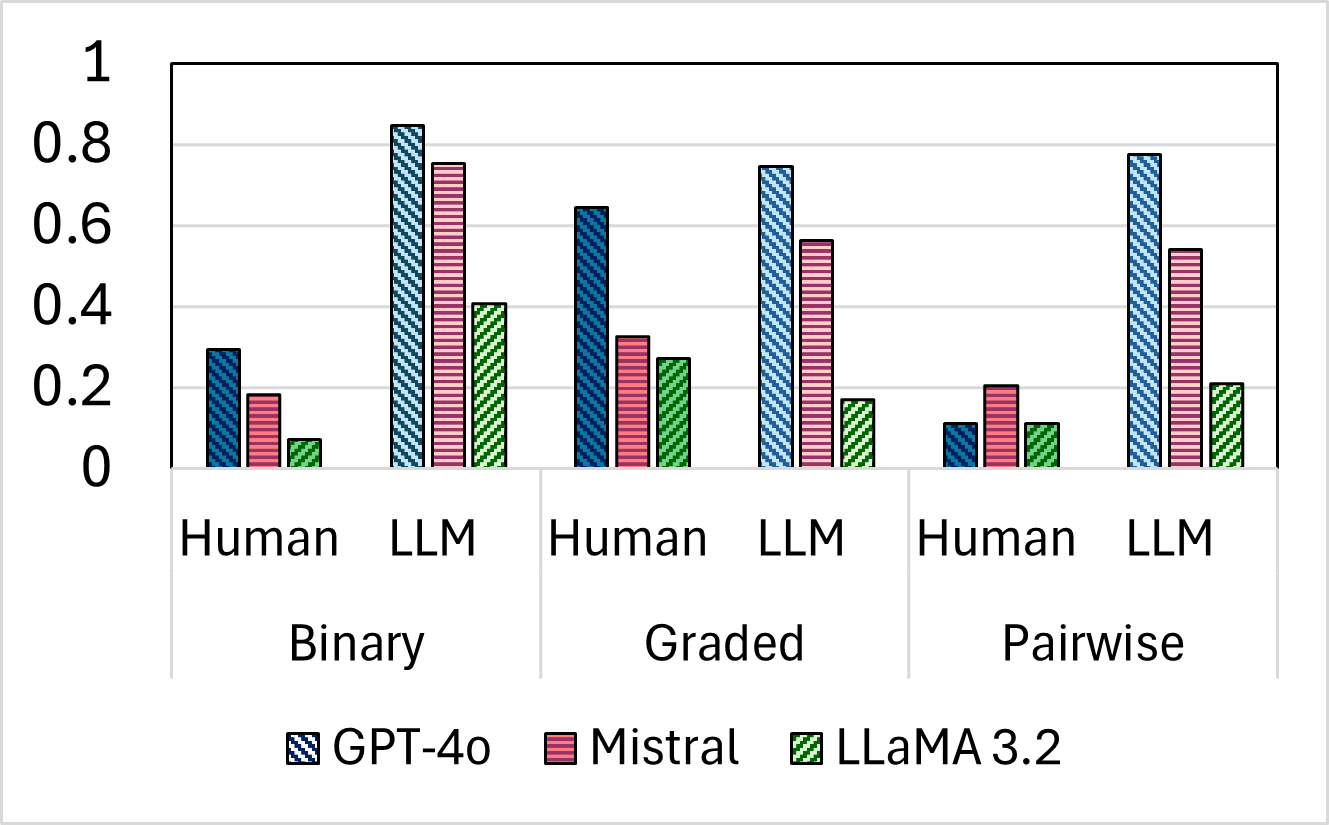}
    \caption{Krippendorff's inter-agreement rate between all the prompts on two datasets.}
    \label{fig:kripendorf}
\end{figure}

\textbf{RQ3. Prompt Robustness Across LLMs:}
Figure~\ref{fig:kripendorf} analyzes inter-agreement rates among different prompt groups using Krippendorff's alpha. Here we measure agreement between different prompt's output, regardless of their alignment with human judgments. The results show that LLM-generated prompts exhibit higher inter-agreement than human-crafted ones, likely due to the greater linguistic diversity in human-generated prompts, as seen in Figure~\ref{fig:prompt_diversity}. This suggests that LLM-generated prompts are more robust than human-crafted ones.
While some human-crafted prompts performed well across all models, prompt effectiveness varies significantly between LLMs, with no single prompt consistently excelling across all models. However, for graded assessments, UMBRELA consistently demonstrated high performance across different LLMs and it emerged as one of the most effective prompts across all models. UMBRELA had previously shown strong correlation with human judgments on TREC DL tracks \cite{upadhyay2024umbrela}. We hypothesize that UMBRELA's strong and consistent performance may stem from how its prompt deconstructs the concept of relevance into finer-grained aspects, such as trustworthiness and alignment with intent. This structured approach likely prevents the LLM from relying on its own interpretation of relevance.

\textbf{RQ4. Model-Specific Sensitivity to Prompts:} 
From Figure~\ref{fig:cohen}, we observe that \texttt{GPT-4o} demonstrates high consistency across most prompts and all relevance assessment approaches. In contrast, the performance of \texttt{LLaMA 3.2} and  \texttt{Mistral} varies significantly depending on the prompt and assessment method. This variability is further confirmed by the variance of agreement reported in Table \ref{tab:agreement_stats}. Notably, \texttt{GPT-4o} exhibits consistently low variance in agreement, particularly when prompted with LLM-crafted prompts.

\section{Conclusion and Limitations}
In this study, we investigated the sensitivity of LLM-based relevance judgments to different prompting strategies across multiple models. We examined how prompts, whether human- or LLM-generated, influence judgment effectiveness, their robustness across different LLMs, and the extent to which models exhibit variability in response to prompt modifications. 
One specific outcome is to confirm the performance of UMBRELA as a leading prompt for LLM-based graded relevance assessment.
Despite these contributions, our study has limitations. Our human participants primarily had a computer science background with experience writing prompts for LLMs. Additionally, we evaluated only three LLMs as judges, limiting the generalizability of our findings.
\bibliographystyle{ACM-Reference-Format}
\balance
\bibliography{sample-base}


\begin{thebibliography}{41}


\ifx \showCODEN    \undefined \def \showCODEN     #1{\unskip}     \fi
\ifx \showISBNx    \undefined \def \showISBNx     #1{\unskip}     \fi
\ifx \showISBNxiii \undefined \def \showISBNxiii  #1{\unskip}     \fi
\ifx \showISSN     \undefined \def \showISSN      #1{\unskip}     \fi
\ifx \showLCCN     \undefined \def \showLCCN      #1{\unskip}     \fi
\ifx \shownote     \undefined \def \shownote      #1{#1}          \fi
\ifx \showarticletitle \undefined \def \showarticletitle #1{#1}   \fi
\ifx \showURL      \undefined \def \showURL       {\relax}        \fi
\providecommand\bibfield[2]{#2}
\providecommand\bibinfo[2]{#2}
\providecommand\natexlab[1]{#1}
\providecommand\showeprint[2][]{arXiv:#2}

\bibitem[Alaofi et~al\mbox{.}(2024)]%
        {alaofi2024generative}
\bibfield{author}{\bibinfo{person}{Marwah Alaofi}, \bibinfo{person}{Negar Arabzadeh}, \bibinfo{person}{Charles~LA Clarke}, {and} \bibinfo{person}{Mark Sanderson}.} \bibinfo{year}{2024}\natexlab{}.
\newblock \showarticletitle{Generative information retrieval evaluation}.
\newblock In \bibinfo{booktitle}{\emph{Information Access in the Era of Generative AI}}. \bibinfo{publisher}{Springer}, \bibinfo{pages}{135--159}.
\newblock


\bibitem[Arabzadeh et~al\mbox{.}(2024a)]%
        {arabzadeh2024adapting}
\bibfield{author}{\bibinfo{person}{Negar Arabzadeh}, \bibinfo{person}{Amin Bigdeli}, {and} \bibinfo{person}{Charles L.~A. Clarke}.} \bibinfo{year}{2024}\natexlab{a}.
\newblock \showarticletitle{Adapting Standard Retrieval Benchmarks to Evaluate Generated Answers}. In \bibinfo{booktitle}{\emph{46th European Conference on Information Retrieval}}. \bibinfo{address}{Glasgow, Scotland}.
\newblock


\bibitem[Arabzadeh and Clarke(2024)]%
        {arabzadeh2024comparison}
\bibfield{author}{\bibinfo{person}{Negar Arabzadeh} {and} \bibinfo{person}{Charles~LA Clarke}.} \bibinfo{year}{2024}\natexlab{}.
\newblock \showarticletitle{A Comparison of Methods for Evaluating Generative IR}.
\newblock \bibinfo{journal}{\emph{arXiv preprint arXiv:2404.04044}} (\bibinfo{year}{2024}).
\newblock


\bibitem[Arabzadeh et~al\mbox{.}(2024b)]%
        {arabzadeh-etal-2024-assessing}
\bibfield{author}{\bibinfo{person}{Negar Arabzadeh}, \bibinfo{person}{Siqing Huo}, \bibinfo{person}{Nikhil Mehta}, \bibinfo{person}{Qingyun Wu}, \bibinfo{person}{Chi Wang}, \bibinfo{person}{Ahmed~Hassan Awadallah}, \bibinfo{person}{Charles L.~A. Clarke}, {and} \bibinfo{person}{Julia Kiseleva}.} \bibinfo{year}{2024}\natexlab{b}.
\newblock \showarticletitle{Assessing and Verifying Task Utility in {LLM}-Powered Applications}. In \bibinfo{booktitle}{\emph{Proceedings of the 2024 Conference on Empirical Methods in Natural Language Processing}}, \bibfield{editor}{\bibinfo{person}{Yaser Al-Onaizan}, \bibinfo{person}{Mohit Bansal}, {and} \bibinfo{person}{Yun-Nung Chen}} (Eds.). \bibinfo{publisher}{Association for Computational Linguistics}, \bibinfo{address}{Miami, Florida, USA}, \bibinfo{pages}{21868--21888}.
\newblock
\href{https://doi.org/10.18653/v1/2024.emnlp-main.1219}{doi:\nolinkurl{10.18653/v1/2024.emnlp-main.1219}}


\bibitem[Arora et~al\mbox{.}(2022)]%
        {arora2022askanythingsimplestrategy}
\bibfield{author}{\bibinfo{person}{Simran Arora}, \bibinfo{person}{Avanika Narayan}, \bibinfo{person}{Mayee~F. Chen}, \bibinfo{person}{Laurel Orr}, \bibinfo{person}{Neel Guha}, \bibinfo{person}{Kush Bhatia}, \bibinfo{person}{Ines Chami}, \bibinfo{person}{Frederic Sala}, {and} \bibinfo{person}{Christopher Ré}.} \bibinfo{year}{2022}\natexlab{}.
\newblock \bibinfo{title}{Ask Me Anything: A simple strategy for prompting language models}.
\newblock
\showeprint[arxiv]{2210.02441}~[cs.CL]
\urldef\tempurl%
\url{https://arxiv.org/abs/2210.02441}
\showURL{%
\tempurl}


\bibitem[Azzopardi et~al\mbox{.}(2024)]%
        {azzopardi2024report}
\bibfield{author}{\bibinfo{person}{Leif Azzopardi}, \bibinfo{person}{Charles~LA Clarke}, \bibinfo{person}{Paul Kantor}, \bibinfo{person}{Bhaskar Mitra}, \bibinfo{person}{Johanne~R Trippas}, \bibinfo{person}{Zhaochun Ren}, \bibinfo{person}{Mohammad Aliannejadi}, \bibinfo{person}{Negar Arabzadeh}, \bibinfo{person}{Raman Chandrasekar}, \bibinfo{person}{Maarten de Rijke}, {et~al\mbox{.}}} \bibinfo{year}{2024}\natexlab{}.
\newblock \showarticletitle{Report on The Search Futures Workshop at ECIR 2024}. In \bibinfo{booktitle}{\emph{ACM SIGIR Forum}}, Vol.~\bibinfo{volume}{58}. ACM New York, NY, USA, \bibinfo{pages}{1--41}.
\newblock


\bibitem[Buckley and Voorhees(2004)]%
        {buckley2004retrieval}
\bibfield{author}{\bibinfo{person}{Chris Buckley} {and} \bibinfo{person}{Ellen~M Voorhees}.} \bibinfo{year}{2004}\natexlab{}.
\newblock \showarticletitle{Retrieval evaluation with incomplete information}. In \bibinfo{booktitle}{\emph{Proceedings of the 27th annual international ACM SIGIR conference on Research and development in information retrieval}}. \bibinfo{pages}{25--32}.
\newblock


\bibitem[Carterette et~al\mbox{.}(2008)]%
        {cbcd08}
\bibfield{author}{\bibinfo{person}{Ben Carterette}, \bibinfo{person}{Paul~N. Bennett}, \bibinfo{person}{David~Maxwell Chickering}, {and} \bibinfo{person}{Susan~T. Dumais}.} \bibinfo{year}{2008}\natexlab{}.
\newblock \bibinfo{booktitle}{\emph{Here or there: {P}reference Judgments for Relevance}}.
\newblock \bibinfo{type}{Computer Science Department Faculty Publication Series}~46. \bibinfo{institution}{University of Massachusetts Amherst}.
\newblock


\bibitem[Chang et~al\mbox{.}(2024)]%
        {chang2024survey}
\bibfield{author}{\bibinfo{person}{Yupeng Chang}, \bibinfo{person}{Xu Wang}, \bibinfo{person}{Jindong Wang}, \bibinfo{person}{Yuan Wu}, \bibinfo{person}{Linyi Yang}, \bibinfo{person}{Kaijie Zhu}, \bibinfo{person}{Hao Chen}, \bibinfo{person}{Xiaoyuan Yi}, \bibinfo{person}{Cunxiang Wang}, \bibinfo{person}{Yidong Wang}, {et~al\mbox{.}}} \bibinfo{year}{2024}\natexlab{}.
\newblock \showarticletitle{A survey on evaluation of large language models}.
\newblock \bibinfo{journal}{\emph{ACM Transactions on Intelligent Systems and Technology}} \bibinfo{volume}{15}, \bibinfo{number}{3} (\bibinfo{year}{2024}), \bibinfo{pages}{1--45}.
\newblock


\bibitem[Chatterjee et~al\mbox{.}(2024)]%
        {chatterjee2024posix}
\bibfield{author}{\bibinfo{person}{Anwoy Chatterjee}, \bibinfo{person}{HSVNS~Kowndinya Renduchintala}, \bibinfo{person}{Sumit Bhatia}, {and} \bibinfo{person}{Tanmoy Chakraborty}.} \bibinfo{year}{2024}\natexlab{}.
\newblock \showarticletitle{POSIX: A Prompt Sensitivity Index For Large Language Models}.
\newblock \bibinfo{journal}{\emph{arXiv preprint arXiv:2410.02185}} (\bibinfo{year}{2024}).
\newblock


\bibitem[Chiang and Lee(2023)]%
        {chiang2023can}
\bibfield{author}{\bibinfo{person}{Cheng-Han Chiang} {and} \bibinfo{person}{Hung-yi Lee}.} \bibinfo{year}{2023}\natexlab{}.
\newblock \showarticletitle{Can large language models be an alternative to human evaluations?}
\newblock \bibinfo{journal}{\emph{arXiv preprint arXiv:2305.01937}} (\bibinfo{year}{2023}).
\newblock


\bibitem[Chiang et~al\mbox{.}(2024)]%
        {chiang2024chatbot}
\bibfield{author}{\bibinfo{person}{Wei-Lin Chiang}, \bibinfo{person}{Lianmin Zheng}, \bibinfo{person}{Ying Sheng}, \bibinfo{person}{Anastasios~Nikolas Angelopoulos}, \bibinfo{person}{Tianle Li}, \bibinfo{person}{Dacheng Li}, \bibinfo{person}{Hao Zhang}, \bibinfo{person}{Banghua Zhu}, \bibinfo{person}{Michael Jordan}, \bibinfo{person}{Joseph~E Gonzalez}, {et~al\mbox{.}}} \bibinfo{year}{2024}\natexlab{}.
\newblock \showarticletitle{Chatbot arena: An open platform for evaluating llms by human preference}.
\newblock \bibinfo{journal}{\emph{arXiv preprint arXiv:2403.04132}} (\bibinfo{year}{2024}).
\newblock


\bibitem[Clarke et~al\mbox{.}(2021a)]%
        {10.1145/3451161}
\bibfield{author}{\bibinfo{person}{Charles L.~A. Clarke}, \bibinfo{person}{Alexandra Vtyurina}, {and} \bibinfo{person}{Mark~D. Smucker}.} \bibinfo{year}{2021}\natexlab{a}.
\newblock \showarticletitle{Assessing Top-$k$ Preferences}.
\newblock \bibinfo{journal}{\emph{ACM Trans. Inf. Syst.}} \bibinfo{volume}{39}, \bibinfo{number}{3}, Article \bibinfo{articleno}{33} (\bibinfo{date}{may} \bibinfo{year}{2021}), \bibinfo{numpages}{21}~pages.
\newblock
\showISSN{1046-8188}
\href{https://doi.org/10.1145/3451161}{doi:\nolinkurl{10.1145/3451161}}


\bibitem[Clarke et~al\mbox{.}(2021b)]%
        {cvs21}
\bibfield{author}{\bibinfo{person}{Charles L.~A. Clarke}, \bibinfo{person}{Alexandra Vtyurina}, {and} \bibinfo{person}{Mark~D. Smucker}.} \bibinfo{year}{2021}\natexlab{b}.
\newblock \showarticletitle{Assessing top-$k$ preferences}.
\newblock \bibinfo{journal}{\emph{ACM Transactions on Information Systems}} \bibinfo{volume}{39}, \bibinfo{number}{3} (\bibinfo{date}{July} \bibinfo{year}{2021}).
\newblock


\bibitem[Cleverdon(1991)]%
        {cleverdon1991significance}
\bibfield{author}{\bibinfo{person}{Cyril~W Cleverdon}.} \bibinfo{year}{1991}\natexlab{}.
\newblock \showarticletitle{The significance of the Cranfield tests on index languages}. In \bibinfo{booktitle}{\emph{Proceedings of the 14th annual international ACM SIGIR conference on Research and development in information retrieval}}. \bibinfo{pages}{3--12}.
\newblock


\bibitem[Craswell et~al\mbox{.}(2021)]%
        {trecdl2020}
\bibfield{author}{\bibinfo{person}{Nick Craswell}, \bibinfo{person}{Bhaskar Mitra}, \bibinfo{person}{Emine Yilmaz}, {and} \bibinfo{person}{Daniel Campos}.} \bibinfo{year}{2021}\natexlab{}.
\newblock \bibinfo{title}{Overview of the TREC 2020 deep learning track}.
\newblock
\showeprint[arxiv]{2102.07662}~[cs.IR]
\urldef\tempurl%
\url{https://arxiv.org/abs/2102.07662}
\showURL{%
\tempurl}


\bibitem[Craswell et~al\mbox{.}(2022)]%
        {trecdl2021}
\bibfield{author}{\bibinfo{person}{Nick Craswell}, \bibinfo{person}{Bhaskar Mitra}, \bibinfo{person}{Emine Yilmaz}, \bibinfo{person}{Daniel Campos}, {and} \bibinfo{person}{Jimmy Lin}.} \bibinfo{year}{2022}\natexlab{}.
\newblock \showarticletitle{Overview of the TREC 2021 deep learning track}. In \bibinfo{booktitle}{\emph{Text REtrieval Conference (TREC)}}. NIST, \bibinfo{publisher}{TREC}.
\newblock
\urldef\tempurl%
\url{https://www.microsoft.com/en-us/research/publication/overview-of-the-trec-2021-deep-learning-track/}
\showURL{%
\tempurl}


\bibitem[Craswell et~al\mbox{.}(2020)]%
        {trecdl2019}
\bibfield{author}{\bibinfo{person}{Nick Craswell}, \bibinfo{person}{Bhaskar Mitra}, \bibinfo{person}{Emine Yilmaz}, \bibinfo{person}{Daniel Campos}, {and} \bibinfo{person}{Ellen~M Voorhees}.} \bibinfo{year}{2020}\natexlab{}.
\newblock \showarticletitle{Overview of the TREC 2019 deep learning track}.
\newblock \bibinfo{journal}{\emph{arXiv preprint arXiv:2003.07820}} (\bibinfo{year}{2020}).
\newblock


\bibitem[Faggioli et~al\mbox{.}(2023)]%
        {faggioli2023perspectives}
\bibfield{author}{\bibinfo{person}{Guglielmo Faggioli}, \bibinfo{person}{Laura Dietz}, \bibinfo{person}{Charles~LA Clarke}, \bibinfo{person}{Gianluca Demartini}, \bibinfo{person}{Matthias Hagen}, \bibinfo{person}{Claudia Hauff}, \bibinfo{person}{Noriko Kando}, \bibinfo{person}{Evangelos Kanoulas}, \bibinfo{person}{Martin Potthast}, \bibinfo{person}{Benno Stein}, {et~al\mbox{.}}} \bibinfo{year}{2023}\natexlab{}.
\newblock \showarticletitle{Perspectives on large language models for relevance judgment}. In \bibinfo{booktitle}{\emph{Proceedings of the 2023 ACM SIGIR International Conference on Theory of Information Retrieval}}. \bibinfo{pages}{39--50}.
\newblock


\bibitem[Farzi and Dietz(2024)]%
        {farzi2024pencils}
\bibfield{author}{\bibinfo{person}{Naghmeh Farzi} {and} \bibinfo{person}{Laura Dietz}.} \bibinfo{year}{2024}\natexlab{}.
\newblock \showarticletitle{Pencils down! automatic rubric-based evaluation of retrieve/generate systems}. In \bibinfo{booktitle}{\emph{Proceedings of the 2024 ACM SIGIR International Conference on Theory of Information Retrieval}}. \bibinfo{pages}{175--184}.
\newblock


\bibitem[Hawking et~al\mbox{.}(1999)]%
        {hawking1999overview}
\bibfield{author}{\bibinfo{person}{David Hawking}, \bibinfo{person}{Ellen Voorhees}, \bibinfo{person}{Nick Craswell}, \bibinfo{person}{Peter Bailey}, {et~al\mbox{.}}} \bibinfo{year}{1999}\natexlab{}.
\newblock \showarticletitle{Overview of the trec-8 web track}. In \bibinfo{booktitle}{\emph{TREC}}.
\newblock


\bibitem[Kazai et~al\mbox{.}(2013)]%
        {kyct13}
\bibfield{author}{\bibinfo{person}{Gabriella Kazai}, \bibinfo{person}{Emine Yilmaz}, \bibinfo{person}{Nick Craswell}, {and} \bibinfo{person}{S.M.M. Tahaghoghi}.} \bibinfo{year}{2013}\natexlab{}.
\newblock \showarticletitle{User Intent and Assessor Disagreement in Web Search Evaluation}. In \bibinfo{booktitle}{\emph{22nd ACM International Conference on Information and Knowledge Management}}. \bibinfo{address}{San Francisco, California}, \bibinfo{pages}{699--708}.
\newblock


\bibitem[Leidinger et~al\mbox{.}(2023)]%
        {leidinger2023languagepromptinglinguisticproperties}
\bibfield{author}{\bibinfo{person}{Alina Leidinger}, \bibinfo{person}{Robert van Rooij}, {and} \bibinfo{person}{Ekaterina Shutova}.} \bibinfo{year}{2023}\natexlab{}.
\newblock \bibinfo{title}{The language of prompting: What linguistic properties make a prompt successful?}
\newblock
\showeprint[arxiv]{2311.01967}~[cs.CL]
\urldef\tempurl%
\url{https://arxiv.org/abs/2311.01967}
\showURL{%
\tempurl}


\bibitem[Li et~al\mbox{.}(2024)]%
        {li2024generation}
\bibfield{author}{\bibinfo{person}{Dawei Li}, \bibinfo{person}{Bohan Jiang}, \bibinfo{person}{Liangjie Huang}, \bibinfo{person}{Alimohammad Beigi}, \bibinfo{person}{Chengshuai Zhao}, \bibinfo{person}{Zhen Tan}, \bibinfo{person}{Amrita Bhattacharjee}, \bibinfo{person}{Yuxuan Jiang}, \bibinfo{person}{Canyu Chen}, \bibinfo{person}{Tianhao Wu}, {et~al\mbox{.}}} \bibinfo{year}{2024}\natexlab{}.
\newblock \showarticletitle{From Generation to Judgment: Opportunities and Challenges of LLM-as-a-judge}.
\newblock \bibinfo{journal}{\emph{arXiv preprint arXiv:2411.16594}} (\bibinfo{year}{2024}).
\newblock


\bibitem[Lu et~al\mbox{.}(2024)]%
        {lu-etal-2024-prompts}
\bibfield{author}{\bibinfo{person}{Sheng Lu}, \bibinfo{person}{Hendrik Schuff}, {and} \bibinfo{person}{Iryna Gurevych}.} \bibinfo{year}{2024}\natexlab{}.
\newblock \showarticletitle{How are Prompts Different in Terms of Sensitivity?}. In \bibinfo{booktitle}{\emph{Proceedings of the 2024 Conference of the North American Chapter of the Association for Computational Linguistics: Human Language Technologies (Volume 1: Long Papers)}}, \bibfield{editor}{\bibinfo{person}{Kevin Duh}, \bibinfo{person}{Helena Gomez}, {and} \bibinfo{person}{Steven Bethard}} (Eds.). \bibinfo{publisher}{Association for Computational Linguistics}, \bibinfo{address}{Mexico City, Mexico}, \bibinfo{pages}{5833--5856}.
\newblock
\href{https://doi.org/10.18653/v1/2024.naacl-long.325}{doi:\nolinkurl{10.18653/v1/2024.naacl-long.325}}


\bibitem[MacAvaney and Soldaini(2023)]%
        {macavaney2023one}
\bibfield{author}{\bibinfo{person}{Sean MacAvaney} {and} \bibinfo{person}{Luca Soldaini}.} \bibinfo{year}{2023}\natexlab{}.
\newblock \showarticletitle{One-shot labeling for automatic relevance estimation}. In \bibinfo{booktitle}{\emph{Proceedings of the 46th International ACM SIGIR Conference on Research and Development in Information Retrieval}}. \bibinfo{pages}{2230--2235}.
\newblock


\bibitem[Meng et~al\mbox{.}(2024)]%
        {meng2024query}
\bibfield{author}{\bibinfo{person}{Chuan Meng}, \bibinfo{person}{Negar Arabzadeh}, \bibinfo{person}{Arian Askari}, \bibinfo{person}{Mohammad Aliannejadi}, {and} \bibinfo{person}{Maarten de Rijke}.} \bibinfo{year}{2024}\natexlab{}.
\newblock \showarticletitle{Query Performance Prediction using Relevance Judgments Generated by Large Language Models}.
\newblock \bibinfo{journal}{\emph{arXiv preprint arXiv:2404.01012}} (\bibinfo{year}{2024}).
\newblock


\bibitem[Rahmani et~al\mbox{.}(2024a)]%
        {rahmani2024LLM4eval}
\bibfield{author}{\bibinfo{person}{Hossein~A Rahmani}, \bibinfo{person}{Clemencia Siro}, \bibinfo{person}{Mohammad Aliannejadi}, \bibinfo{person}{Nick Craswell}, \bibinfo{person}{Charles~LA Clarke}, \bibinfo{person}{Guglielmo Faggioli}, \bibinfo{person}{Bhaskar Mitra}, \bibinfo{person}{Paul Thomas}, {and} \bibinfo{person}{Emine Yilmaz}.} \bibinfo{year}{2024}\natexlab{a}.
\newblock \showarticletitle{Llm4eval: Large language model for evaluation in ir}. In \bibinfo{booktitle}{\emph{Proceedings of the 47th International ACM SIGIR Conference on Research and Development in Information Retrieval}}. \bibinfo{pages}{3040--3043}.
\newblock


\bibitem[Rahmani et~al\mbox{.}(2024b)]%
        {rahmani2024report1stworkshoplarge}
\bibfield{author}{\bibinfo{person}{Hossein~A. Rahmani}, \bibinfo{person}{Clemencia Siro}, \bibinfo{person}{Mohammad Aliannejadi}, \bibinfo{person}{Nick Craswell}, \bibinfo{person}{Charles L.~A. Clarke}, \bibinfo{person}{Guglielmo Faggioli}, \bibinfo{person}{Bhaskar Mitra}, \bibinfo{person}{Paul Thomas}, {and} \bibinfo{person}{Emine Yilmaz}.} \bibinfo{year}{2024}\natexlab{b}.
\newblock \bibinfo{title}{Report on the 1st Workshop on Large Language Model for Evaluation in Information Retrieval (LLM4Eval 2024) at SIGIR 2024}.
\newblock
\showeprint[arxiv]{2408.05388}~[cs.IR]
\urldef\tempurl%
\url{https://arxiv.org/abs/2408.05388}
\showURL{%
\tempurl}


\bibitem[Razavi et~al\mbox{.}(2025)]%
        {razavi2025benchmarking}
\bibfield{author}{\bibinfo{person}{Amirhossein Razavi}, \bibinfo{person}{Mina Soltangheis}, \bibinfo{person}{Negar Arabzadeh}, \bibinfo{person}{Sara Salamat}, \bibinfo{person}{Morteza Zihayat}, {and} \bibinfo{person}{Ebrahim Bagheri}.} \bibinfo{year}{2025}\natexlab{}.
\newblock \showarticletitle{Benchmarking Prompt Sensitivity in Large Language Models}.
\newblock \bibinfo{journal}{\emph{arXiv preprint arXiv:2502.06065}} (\bibinfo{year}{2025}).
\newblock


\bibitem[Sakai and Zeng(2020)]%
        {sz20}
\bibfield{author}{\bibinfo{person}{Tetsuya Sakai} {and} \bibinfo{person}{Zhaohao Zeng}.} \bibinfo{year}{2020}\natexlab{}.
\newblock \showarticletitle{Good evaluation measures based on document preferences}. In \bibinfo{booktitle}{\emph{43rd International ACM SIGIR Conference on Research and Development in Information Retrieval}}. \bibinfo{pages}{359–368}.
\newblock


\bibitem[Salemi and Zamani(2024)]%
        {salemi2024evaluating}
\bibfield{author}{\bibinfo{person}{Alireza Salemi} {and} \bibinfo{person}{Hamed Zamani}.} \bibinfo{year}{2024}\natexlab{}.
\newblock \showarticletitle{Evaluating retrieval quality in retrieval-augmented generation}. In \bibinfo{booktitle}{\emph{Proceedings of the 47th International ACM SIGIR Conference on Research and Development in Information Retrieval}}. \bibinfo{pages}{2395--2400}.
\newblock


\bibitem[Sander and Dietz(2021)]%
        {sander2021exam}
\bibfield{author}{\bibinfo{person}{David~P Sander} {and} \bibinfo{person}{Laura Dietz}.} \bibinfo{year}{2021}\natexlab{}.
\newblock \showarticletitle{EXAM: How to Evaluate Retrieve-and-Generate Systems for Users Who Do Not (Yet) Know What They Want.}. In \bibinfo{booktitle}{\emph{DESIRES}}. \bibinfo{pages}{136--146}.
\newblock


\bibitem[Sclar et~al\mbox{.}(2023)]%
        {sclar2023quantifying}
\bibfield{author}{\bibinfo{person}{Melanie Sclar}, \bibinfo{person}{Yejin Choi}, \bibinfo{person}{Yulia Tsvetkov}, {and} \bibinfo{person}{Alane Suhr}.} \bibinfo{year}{2023}\natexlab{}.
\newblock \showarticletitle{Quantifying Language Models' Sensitivity to Spurious Features in Prompt Design or: How I learned to start worrying about prompt formatting}.
\newblock \bibinfo{journal}{\emph{arXiv preprint arXiv:2310.11324}} (\bibinfo{year}{2023}).
\newblock


\bibitem[Thomas et~al\mbox{.}(2023)]%
        {thomas2023large}
\bibfield{author}{\bibinfo{person}{Paul Thomas}, \bibinfo{person}{Seth Spielman}, \bibinfo{person}{Nick Craswell}, {and} \bibinfo{person}{Bhaskar Mitra}.} \bibinfo{year}{2023}\natexlab{}.
\newblock \showarticletitle{Large Language Models Can Accurately Predict Searcher Preferences}.
\newblock \bibinfo{journal}{\emph{arXiv preprint arXiv:2309.10621}} (\bibinfo{year}{2023}).
\newblock


\bibitem[Upadhyay et~al\mbox{.}(2024a)]%
        {upadhyay2024largescalestudyrelevanceassessments}
\bibfield{author}{\bibinfo{person}{Shivani Upadhyay}, \bibinfo{person}{Ronak Pradeep}, \bibinfo{person}{Nandan Thakur}, \bibinfo{person}{Daniel Campos}, \bibinfo{person}{Nick Craswell}, \bibinfo{person}{Ian Soboroff}, \bibinfo{person}{Hoa~Trang Dang}, {and} \bibinfo{person}{Jimmy Lin}.} \bibinfo{year}{2024}\natexlab{a}.
\newblock \bibinfo{title}{A Large-Scale Study of Relevance Assessments with Large Language Models: An Initial Look}.
\newblock
\showeprint[arxiv]{2411.08275}~[cs.IR]
\urldef\tempurl%
\url{https://arxiv.org/abs/2411.08275}
\showURL{%
\tempurl}


\bibitem[Upadhyay et~al\mbox{.}(2024b)]%
        {upadhyay2024umbrela}
\bibfield{author}{\bibinfo{person}{Shivani Upadhyay}, \bibinfo{person}{Ronak Pradeep}, \bibinfo{person}{Nandan Thakur}, \bibinfo{person}{Nick Craswell}, {and} \bibinfo{person}{Jimmy Lin}.} \bibinfo{year}{2024}\natexlab{b}.
\newblock \showarticletitle{UMBRELA: UMbrela is the (Open-Source Reproduction of the) Bing RELevance Assessor}.
\newblock \bibinfo{journal}{\emph{arXiv preprint arXiv:2406.06519}} (\bibinfo{year}{2024}).
\newblock


\bibitem[Voorhees(2000)]%
        {voorhees2000report}
\bibfield{author}{\bibinfo{person}{Ellen~M Voorhees}.} \bibinfo{year}{2000}\natexlab{}.
\newblock \showarticletitle{Report on trec-9}. In \bibinfo{booktitle}{\emph{ACM SIGIR Forum}}, Vol.~\bibinfo{volume}{34}. ACM New York, NY, USA, \bibinfo{pages}{1--8}.
\newblock


\bibitem[Xie et~al\mbox{.}(2020)]%
        {xie20}
\bibfield{author}{\bibinfo{person}{Xiaohui Xie}, \bibinfo{person}{Jiaxin Mao}, \bibinfo{person}{Yiqun Liu}, \bibinfo{person}{Maarten de Rijke}, \bibinfo{person}{Haitian Chen}, \bibinfo{person}{Min Zhang}, {and} \bibinfo{person}{Shaoping Ma}.} \bibinfo{year}{2020}\natexlab{}.
\newblock \showarticletitle{Preference-based evaluation metrics for web image search}. In \bibinfo{booktitle}{\emph{43st Annual International ACM SIGIR Conference on Research and Development in Information Retrieval}}. \bibinfo{address}{Xi'an, China}.
\newblock


\bibitem[Yan et~al\mbox{.}(2022)]%
        {Yan_2022}
\bibfield{author}{\bibinfo{person}{Xinyi Yan}, \bibinfo{person}{Chengxi Luo}, \bibinfo{person}{Charles L.~A. Clarke}, \bibinfo{person}{Nick Craswell}, \bibinfo{person}{Ellen~M. Voorhees}, {and} \bibinfo{person}{Pablo Castells}.} \bibinfo{year}{2022}\natexlab{}.
\newblock \showarticletitle{Human Preferences as Dueling Bandits}. In \bibinfo{booktitle}{\emph{Proceedings of the 45th International ACM SIGIR Conference on Research and Development in Information Retrieval}} \emph{(\bibinfo{series}{SIGIR ’22})}. \bibinfo{publisher}{ACM}.
\newblock
\href{https://doi.org/10.1145/3477495.3531991}{doi:\nolinkurl{10.1145/3477495.3531991}}


\bibitem[Zhuang et~al\mbox{.}(2023)]%
        {zhuang2023beyond}
\bibfield{author}{\bibinfo{person}{Honglei Zhuang}, \bibinfo{person}{Zhen Qin}, \bibinfo{person}{Kai Hui}, \bibinfo{person}{Junru Wu}, \bibinfo{person}{Le Yan}, \bibinfo{person}{Xuanhui Wang}, {and} \bibinfo{person}{Michael Berdersky}.} \bibinfo{year}{2023}\natexlab{}.
\newblock \showarticletitle{Beyond Yes and No: Improving Zero-Shot LLM Rankers via Scoring Fine-Grained Relevance Labels}.
\newblock \bibinfo{journal}{\emph{arXiv preprint arXiv:2310.14122}} (\bibinfo{year}{2023}).
\newblock


\end{thebibliography}
\balance
\end{document}